\documentclass[acmsmall, nonacm]{acmart}

\usepackage{comment}
\usepackage{xcolor}

\newcommand{\ie}{\textit{i}.\textit{e}.}

\newcommand{\etc}{\textit{etc}. }

\AtBeginDocument{%
  \providecommand\BibTeX{{%
    \normalfont B\kern-0.5em{\scshape i\kern-0.25em b}\kern-0.8em\TeX}}}

\setcopyright{rightsretained}
\copyrightyear{2021}
\acmYear{2021}
\acmDOI{}

\acmJournal{TOPS}
\acmVolume{1}
\acmNumber{1}
\acmArticle{1}
\acmMonth{7}

\begin{document}

\title{Encrypted Data Processing}

\author{Jessica Tseng}
\email{jhtseng@us.ibm.com}
\author{Gianfranco Bilardi}
\authornote{The author is also affiliated with University of Padova, Italy (\textit{bilardi@dei.unipd.it}).}
\email{gbilardi@us.ibm.com}
\author{Kattamuri Ekanadham}
\email{eknath@us.ibm.com}
\author{Manoj Kumar}
\email{manoj1@us.ibm.com}
\author{Jose Moreira}
\email{jmoreira@us.ibm.com}
\author{P. C. Pattnaik}
\email{pratap@us.ibm.com}
\affiliation{%
  \institution{IBM Research Division}
  \streetaddress{P.O. Box 218}
  \city{Yorktown Heights}
  \state{NY}
  \country{USA}
  \postcode{10598}
}

\renewcommand{\shortauthors}{Tseng, et al.}

\begin{abstract}
In this paper, we present a comprehensive architecture for confidential computing, which we show to be general purpose and quite efficient. It executes the application as is, without any added burden or discipline requirements from the application developers. Furthermore, it does not require the trust of system software at the computing server and does not impose any added burden on the communication subsystem.  The proposed \textbf{\textit{Encrypted Data Processing (EDAP)}} architecture accomplishes confidentiality, authenticity, and freshness of the key-based cryptographic data protection by adopting data encryption with a multi-level key protection scheme.  It guarantees that the user data is visible only in non-privileged mode to a designated program trusted by the data owner on a designated hardware, thus protecting the data from an untrusted hardware, hypervisor, OS, or other users’ applications.  The cryptographic keys and protocols used for achieving these confidential computing requirements are described in a use case example.  Encrypting and decrypting data in an EDAP-enabled processor can lead to performance degradation as it adds cycle time to the overall execution.  However, our simulation result shows that the slowdown is only 6\% on average across a collection of commercial workloads when the data encryption engine is placed between the L1 and L2 cache.  We demonstrate that the EDAP architecture is valuable and practicable in the modern cloud environment for confidential computing.  EDAP delivers a zero trust model of computing where the user software does not trust system software and vice versa. 
\end{abstract}

\maketitle

\section{Introduction and Background}
In recent years, cloud computing has become the ubiquitous platform for server portion of the computing.
There have been many definitions of cloud computing, more or less implying the same attributes.
Here we quote the definition from NIST (National Institute of Standards and Technology), that is
{\sl the Cloud computing is a model for enabling ubiquitous, convenient, on-demand network access
to a shared pool of configurable computing resources that can be rapidly provisioned and
released with minimal management effort or service provider interaction.} \cite{NIST_Cloud}.

To facilitate the use of cloud computing by data owners (\textbf{DO}s)
to get the desired computations and analysis done
on their sensitive data, while preserving the benefit of cloud computing,
we propose a comprehensive architecture for confidential computing.
This architecture, Encrypted Data Processing (\textbf{EDAP}), is general purpose and quite efficient. 
It executes the application as is, without  any added burden or discipline required from the application developers. Furthermore, it does not require the trust of system software at the cloud servers, and does not impose any added  burden on the communication subsystem.
EDAP achieves its goals by reducing the trusted hardware/software footprint in the cloud server.

The Encrypted Data Processing (EDAP) architecture provides security for data-in-use, augmenting the traditionally available mechanisms for securing data-at-rest and data-in-transmission.
Even though the cloud is performing the computations for the DO, the data owner's cleartext (unencrypyted) data is not visible to any software other than the DO's thread in problem (non-privileged) state.  
The hypervisor, VMs, OS, or the service software like dump  {\etc}, in the cloud, including the hardware thread processing the data owner's data, when in elevated privilege mode, can not access the data owner's cleartext data.  This denial of access is maintained even in presence of security vulnerabilities in the entire software stack, except in the DO's program, and in the presence of insider threats from rogue system administrators.

This trusted hardware footprint could be a processor core with its L1 cache, the processor core without the L1 cache, or individual functional units of the processor core.  Data stays encrypted outside the trusted hardware footprint, it is decrypted as it enters the trusted footprint, and results of computation are re-encrypted as  they leave the trusted footprint.  The three design points offer different trade-offs between the size of the secure footprint and the extent of reuse of cleartext data.  The DO exchanges data and program with the trusted footprint over an encrypted channel, using a processor specific public-private key pair, where the private key is burnt into the hardware and assumed to be inaccessible outside the trusted footprint. 
In this paper we will present further details of the hardware extensions in the processor's trusted footprint needed to support the cryptographic operations for each of the three design points, needed to
protect the data owner's data while exploiting the reuse of cleartext from the L1 cache, register file(s), or cleartext buffers in the individual functional units, respectively.  The rest of the components in the cloud are thus reduced to be pure transporter without access to DO's cleartext data. 
 
To reduce the trusted software footprint, particularly the reliance on security of
address translation capabilities in firmware, hypervisor, or OS, EDAP ensures that cleartext data in the trusted hardware footprint can only be accessed by an instruction from the program specified by the DO, i.e., fetched from the effective address of the instruction in the specified program. The cryptographic protocol support needed from the trusted hardware to enforce this requirement is also discussed.  This ensures that instructions of any program, other than the one authorized by the DO running in the problem state, can not access DO's cleartext data by exploiting vulnerabilities in address translation mechanisms of firmware, hypervisor, or OS.

Conceptually the EDAP approach is akin to Fully Homomorphic Encryption (FHE) computing in intent.  
FHE can be seen as the limiting case of the the three design points presented in this paper, where the 
trusted hardware footprint has been reduced to nothing!
EDAP significantly reduces the performance overhead of achieving a degree of confidentiality similar to that of FHE by utilizing the micro-architecture 
of the modern out-of-order processors, and appropriately augmenting them.

The decryption of DO's data, operation on it, and re-encryption of the result, can be
made an architecturally atomic operation, ensuring that data never exists in the system in decrypted form \cite{LaurenBiernacki}.
However, the decryption and encryption overhead would significantly slow down the overall performance.  
The three extensions proposed by us make progressively greater reuse of the cleartext (unencrypted) data by increasing the size of the trusted hardware footprint.  In this paper, we present simulation results for our easiest to implement confidential computing model, i.e., confining cleartext to the processor core and its L1 cache.  The simulations show that the overhead of this proposed approach is a modest 6\% across commercial workload, on average.

\subsection{Prior art}
Intel Software Guard Extensions SGX \cite{ShayGueron} too guarantees that access to DO's program and data is restricted to DO's program and provides for confidentiality, integrity, and freshness (safeguarding from replay attacks).
The trusted hardware footprint in SGX is the entire processor chip, including the last level cache.
Two key elements of SGX are: 1) Trusted firmware that carves out a region of memory for use only by DO's application and the firmware, excluding system software access to this region, 2) encrypting each cache line and then hashing it along with a nonce (write-counter) to generate a MAC (Message Authentication Code) which is verified in the processor every time the cache line is use.  The main draw back of the SGX approach is that the nonce have to be protected, and that entails the overhead maintaining a Merkle tree of the nonce values in memory and caching them on chip for performance.

Williams and Boivie in \cite{Boivie2} describe a solution similar to SGX in placement of encryption and decryption happening between the last level cache and main memory.  They encode the session identifier of a program under execution with each encrypted cache line to maintain the integrity of a users application and data, an element of our solution too.  Replay attacks are handled in a manner similar to SGX, by maintaining an integrity tree.  
Their approach is similar to ours in the sense that trusted software footprint is minimized by excluding OS and hypervisor etc. from it.  

The main contributions of this paper are:
\begin{enumerate}
\item
Three design points for trusted hardware footprint that enable reuse of cleartext data for confidential computing.
Extensions needed in the processor, or more precisely the trusted footprint, to reuse the cleartext data
without exposing it to any software other than DOs program running in problem (non-privelaged) state.
These hardware extensions are covered in Section \ref{Implementation}.
\item
The cryptographic protocol to: 1) enable a DO to communicate securely with the trusted hardware footprint, ensuring that the data exists in cleartext only on the trusted footprint; and 2) enforce that DO's cleartext data is accessed only by instructions of a designated program fetched from its designated effective address. The protocol is presented in Section \ref{EDAPArchitecture}. 
\item
A cycle accurate simulation for an out-of-order pipelined processor, a POWER10 like processor. We used a modified version of the cycle-accurate simulator used for the design of POWER processors, which accounts for the encryption/decryption delays.
\end{enumerate}
  
The next section introduces the actors, the trust model, the cryptographic objects and the cryptographic protocols to establish the context for the subsequent sections described above.  Section \ref{UsageScenarios} illustrates the secure computation protocol implemented using the EDAP capabilities.  Section \ref{Sec:PerformanceEvaluation} discusses the simulation model and its projected performance results.  Finally, Section \ref{sec:Conclusions} presents the conclusions of the paper, the main message being that for, an aggressively out of order processor, the overhead of not trusting the software is quite smaller than one may naïvely expect and may all in the realm of acceptable cost.

\section{Data Confidentiality Preservation Protocol (DCPP)}
\label{Protocol} 
In this section we describe the core security related aspects of EDAP, namely the principals, the threat model addressed, cryptographic objects they use, and the cryptographic protocol.

\subsection{Principals in EDAP}
\label{EDAP_Principals}

\subsubsection{The data owner(s) \textbf{DO}}
Several owners of data can be providing data for a computation performed on behalf of a compute customer.  Any subset of them would opt to keep their data encrypted in the computations being performed on their data.  To keep the discussion simple, for now, we will assume that there is only one data owner, and she is also the compute customer.  We will also assume that the DO is interested in: 1) keeping her entire data confidential; and (2) ensuring that her data is accessible as cleartext only to the application designated by her.

\subsubsection{The application author \textbf{AA}}
There could be multiple application authors providing the application or libraries that will be statically linked to the application.  Once again, to keep the discussion simple, we will assume a single AA providing the statically linked application.

\subsubsection{The platform provider \textbf{PP}}
Provides the virtual machines or cloud infrastructure for the computations.
The PP is not trusted with confidentiality of DO's data, he can be an adversary or his lax security could provide adversaries access to the DO's data.

\subsection{The Threat Model}
\label{threatModel}

There can be multiple malicious actors, including the PP himself and external actors, however we can model the situation as a single malicious actor representing the following combined capabilities, \ie , threat model: 
\begin{itemize}
\item
The adversary can gain privileged access to PP's systems, and thus gain access to DO's data in the memory, L2 and stages of memory hierarchy in between. Data in L1 caches is assumed safe and can be in clear-text, when L1 is in the 
trusted software footprint.
\item
The adversary can exploit vulnerabilities in PP's system software (OS and hypervisor) or inject malicious code into system the software stack to get access to DO's data.
\item
The adversary can manipulate AA's code, unless the code is secured by cryptography.  This allows him to cast DO's encrypted data into the clear, and access it. Code is secured cryptographically by ensuring that: 1) it can
be executed only by a designated thread-id, and from its intended location in thread's address space.
\end{itemize}

The goal of the paper is to provide cryptographic capabilities to mitigate the threats posed by the untrusted platform provider or malicious actors using the the cryptographic objects and protocols summarized in rest of this section and detailed in Section \ref{UsageScenarios}.  The protocols rely on the EDAP architecture and hardware enhancements described in Sections \ref{EDAPArchitecture} and \ref{Implementation}.

\subsection{Key cryptography objects}

\begin{figure}
  \includegraphics[scale=.4]{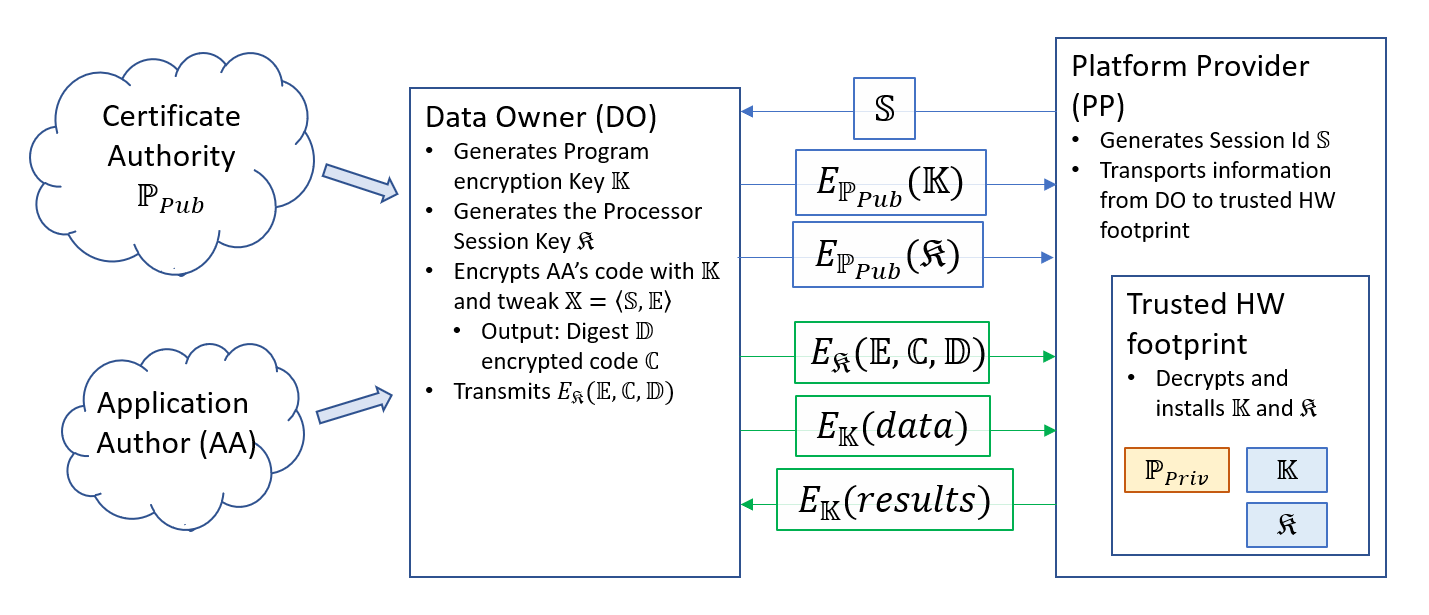}
  \caption{EDAP Protocol for Confidential Computing}
  \label{fig:Protocol}
\end{figure}

\begin{enumerate}
\item
A processor private-public key pair $\langle {\mathbb{P}}_{Priv}, 
{\mathbb{P}}_{Pub} \rangle$.  The private key ${\mathbb{P}}_{Priv}$ is burnt into the trusted hardware footprint as shown in Figure \ref{fig:Protocol}.  This key is fixed for the life time of the hardware.
\item
A secure processing session identifier (\textbf{SEID}) $\mathbb{S}$.  Generated by the PP, for the duration of executing the program provided by the DO along with its data.
If these properties are not satisfied, the DO's program will not work, but her data will not be compromised.
\item
A program encryption key $\mathbb{K}$.  A symmetric key generated by the DO to encrypt her program before transmitting it to PP in the step below. (Possibly using AES-XTS). This encrypted program $C {\mathbb{K}}$, is stored in the memory in its encrypted form.
\item
Processor session key $\mathfrak{K}$. A symmetric key generated by the DO to encrypt her program, a second time after encryption by $\mathbb{K}$, for transmitting it over a secure channel (possibly using AES-GCM).
AES-XTS provides block-wise random access to encrypted code.  The second encryption (AES-GCM)is used for by the trusted footprint to authenticate the author of the code being executed and to prevent replay of DO's program. This doubly encrypted code is the code resident in program memory.
\end{enumerate}

The secure computing protocol is defined in detail in Section \ref{UsageScenarios}.  Its key steps are summarized in Figure \ref{fig:Protocol}
to illustrate the use of the cryptographic objects and provide the context for the next two sections.

\subsection{The root of trust}
\label{rootTrust}
The hardware root of trust is the secure hardware footprint in the EDAP enabled processor, i.e., the encryption logic within the confines of the processor core and its L1 cache.  It consists of the conventional processor core and following additional cryptographic elements and capabilities:
\begin{enumerate}
\item
The burnt in Processor Private Key ${\mathbb{P}}_{Priv}$, which never leaves the processor core. It also stores a firmware/hypervisor 'updateable' Processor Public key ${\mathbb{P}}_{Pub}$, certified by a certification issuing authority.  Public Key is updateable because we may need to change the certification authority during the lifetime of the Processor.
\item
Registers to hold the session identifier SEID, the program encryption key
$\mathbb{K}$, and the session key $\mathfrak{K}$.
\item
A microcontroller built into the processor core, the trusted hardware footprint, to carry out the tasks of installing the cryptographic keys in their registers and carrying out the necessary decryption functions.
\end{enumerate}

Outside the above trusted processor core, rest of the hardware is assumed to be untrusted.  The trusted software footprint is
the code signed by the DO.  OS and hypervisor code is untrusted.

\section{EDAP Architecture} 
\label{EDAPArchitecture} 
EDAP architecture achieves secure execution of processes,
particularly confidentiality of data,
by storing encrypted data in memory and implementing a multi-level key protection scheme.  The encrypted symmetric keys for memory encryption are exchanged between the data owner and the trusted processor over a channel using the processor's public-key/private-key pair, preventing the platform provider or other applications to access them.  Moreover, the memory encryption keys are stored in 
encrypted state in the processor, encrypted by a processor die specific public key (its provate key counterpart burnt into the processor), 
to prevent unauthorized access.  Thus, the EDAP protected data is only visible to the intended processor that is authorized by the data owner to process the data.  

Additionally, the EDAP architecture guarantees that only the authorized program running on the designated EDAP enabled processor can decrypt data-owner's confidential data.  Furthermore, the instructions of the authorized program can only be executed when they are issued from the program-memory address that is specified in the signed program, preventing unauthorized accesses from being mapped to cleartext by erroneous or malicious privileged code.


EDAP is built on three fundamental assumptions.  First, EDAP enabled machines are trusted and cannot be sabotaged physically.  Next, the cryptographic technology is robust enough and the cleartext data can be obtained only when the encryption keys are known.  Lastly, the program authorized by the data owner is not malicious and its certification is trusted.  Base on these assumptions, there are five major elements in the EDAP architecture that safeguard data owner’s data.  Section \ref{keypair}, Section \ref{transfercontrol} and Section \ref{restrictionsOnCode} introduce the security features in EDAP enabled processors that provides the exclusive data encryption and data protection.  The secure protocols to exchange data between the data owner and the trusted processor through an intermediary, the untrusted platform provider, is described in sections \ref{opaque} and \ref{sharingdata}.

\subsection{Public/Private Key Pair}
\label{keypair}
The processor manufacturer assigns a unique pair of public/private key to each EDAP enabled machine for public-key cryptography.  A message that is encrypted with the public key can only be decrypted by the paired private key.  The public key is published by the manufacturer and it can be verified by users.  The unclonable private key is burned into the EDAP enabled processor hardware in the factory before its packaging and is not readable outside the trusted hardware footprint.  
Since the private key is unique to each processor and an encrypted message can only be decrypted with the matching private key, message that is encrypted with a particular public key can only be decrypted on the processor with the paired private key.  This arrangement ensures both the data confidentiality and machine authentication in EDAP.  

\begin{figure}
  \includegraphics[scale=.4]{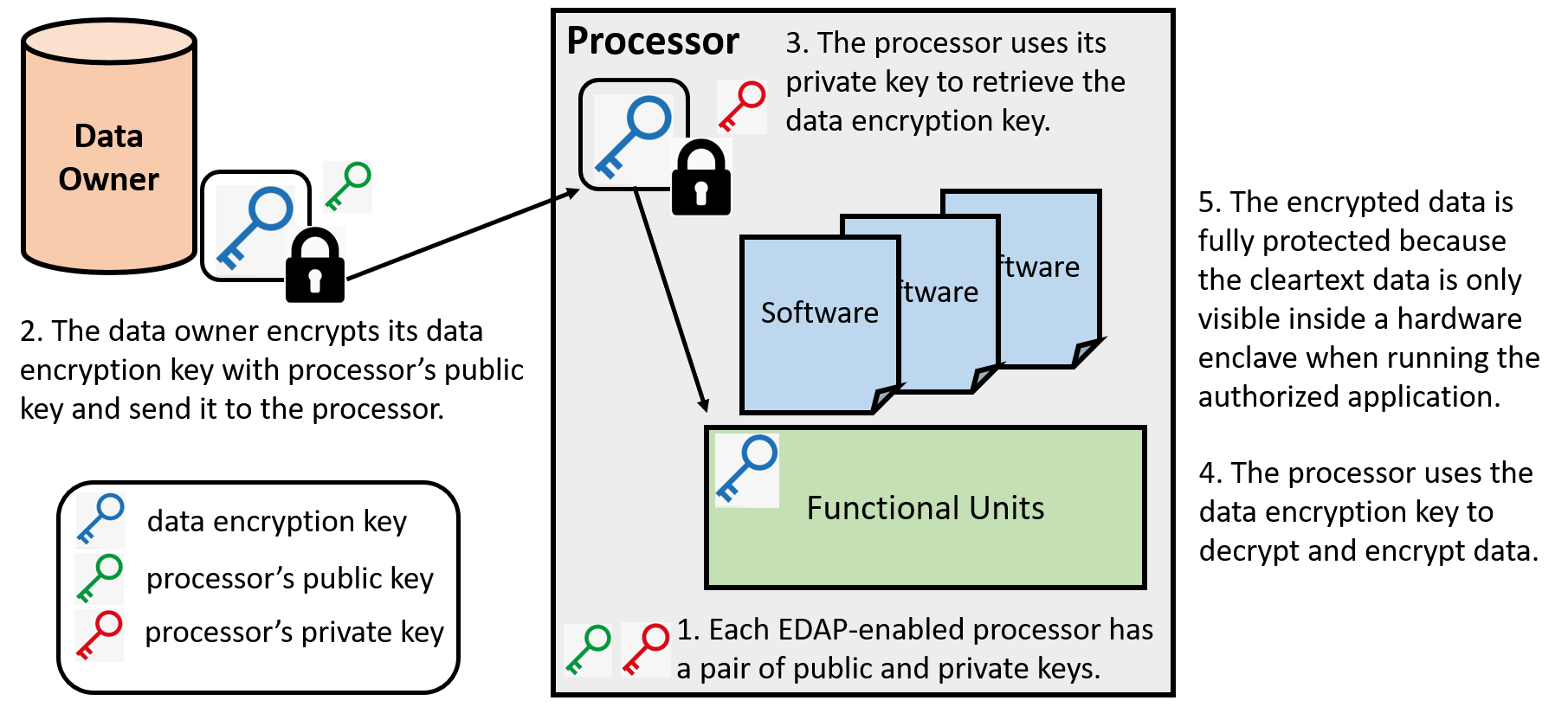}
  \caption{EDAP Encryption Key Management}
  \label{fig:PublicPrivateKeyPair}
\end{figure}

EDAP architecture uses the above mentioned public/private key pair to wrap the data encryption keys in its multi-level key protection scheme as shown in Figure \ref{fig:PublicPrivateKeyPair}.  Data owner first encrypts the data with data encryption keys.  Next, the data encryption keys are encrypted with the public key of the authorized EDAP enabled machine before being send to its platform provider.   Outside the trusted hardware footprint, the data encryption keys are stored in the encrypted form and can only be decrypted by the authorized processor withing the trusted footprint to further decrypt the protected data for processing.
The cleartext data encryption keys are not readable outside the trusted hardware footprint.
The authorized processor then encrypts the results with the encryption key of data owner before storing it back into the memory to send back to the data owner.    

\subsection{Transferring of Control} 
\label{transfercontrol}
EDAP enabled processors take additional steps when transferring the control to guarantee cleartext data is only visible to the problem-state of the authorized application.  The processor clears all cleartext data along with its derived meta data and disengages the encryption engine before any transfer of control into supervisor/hypervisor.  Similarly, any transfer of control back to problem state of the authorized application requires clearing spaces that hold cleartext data and reengaging encryption engine.  Depending on the implementation, supporting instructions could be added to perform these operations and to save and to restore program state without exposing the cleartext data.  For example, privilege instructions would be required to restore and to save thread states without exposing the data to hypervisor/supervisor for context switch if any of processor states are in cleartext. 

\subsection{Restrictions on Code that Accesses Protected Data} 
\label{restrictionsOnCode}
In addition to direct access to DO's cleartext data by unauthorized programs, confidentiality of the data can also be compromised by allowing unauthorized replays of the DO authorized program, or rearranging the instructions in the DO's program.  This threat is averted by having the DO sign the program authorized by him with the address of the code block, and the unique thread-id provide by the platform provider. 

\subsection{Opaque Transaction on Untrusted Platform Provider}
\label{opaque}
As the data owner cannot trust the platform provider, all the transactions from the data owner to the platform provider are opaque in EDAP architecture because they are all encrypted with the public key of the authorized EDAP enabled processor and only the intended processor can decrypt the program and data exchanged with the DO for processing.    Confidential transactions such as session keys, data encryption keys and program executables are not visible to the platform provider and other software.  This provides a secure end-to-end connection between the data owner and the authorized and authenticated trusted hardware footprint in the EDAP enabled machine without placing trust on the platform provider.  Additionally, the certified executable from the data owner includes all necessary problem-state code (i.e. the secure executable is statically linked) since none of other software can be trusted, including installed libraries.  On the other hand, it is safe to use supervisor/hypervisor service that causes a change of privilege because necessary steps are taken to protect the data during transferring of control as described in Section \ref{transfercontrol}.    

\subsection{Sharing Additional Protected Data}
\label{sharingdata}
Protected data can be part of the secure executable as static data.  Once the authorized application is running on an EDAP enabled machine, there are several approaches to share additional secret data.  One way is to establish a secure socket-layer between the data owner and the authorized application.  This is a direct connection instead of the secure session through the platform provider.  SSL credential can be built into the executable, independent of the memory encryption scheme to provide additional level of security.  Another way to share additional protected data is to place it in a shared storage and its keys can be built into the executable or loaded at run time.


\section{EDAP Implementation}
\label{Implementation} 
Implementation and performance variation of EDAP architecture depend on where we place the data encryption engine.  The encryption engine converts data from cyphertext to cleartext and from cleartext to cyphertext.  It is an essential part of EDAP architecture to control read accesses to the protected data.  The placement of encryption engine determines the number of times a piece of data is decrypted and re-encrypted.  Decrypting and encrypting data adds cycle time to the overall execution and can significantly decrease the overall performance.  Moreover, depending on the placement, the protected data is presented as cleartext throughout different part of processor core and requires different level of processor support to purge them when transferring control.

\subsection{Inside Functional Unit Enclave}
\begin{figure}
  \includegraphics[scale=.5]{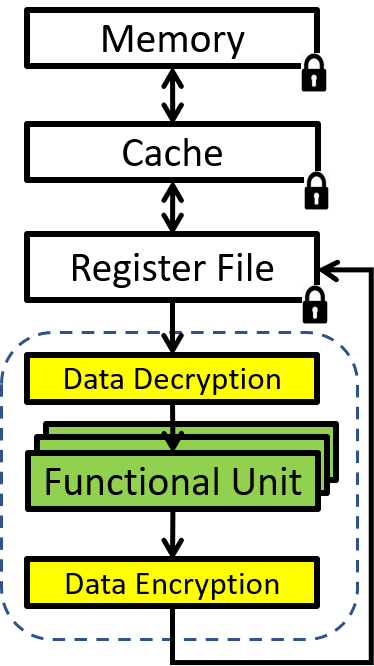}
  \caption{EDAP implementation with data encryption engine placed inside the function unit enclave}
  \label{fig:EDAPimplementation1}
\end{figure}

The most conservative implementation of EDAP architecture is placing the data encryption engine inside the function unit enclave to keep the data encrypted throughout the processor.  The data is decrypted once it enters the functional unit for execution and the computed result is encrypted before being written back to the register file as shown in Figure \ref{fig:EDAPimplementation1}.  This adversely impacts the latency of instruction execution overall performance of the application.  However, the context switch can be performed normally as the entire processor state remains encrypted and is not visible without the correct authorization to be decrypted in the functional unit.

\begin{figure}
  \includegraphics[scale=.5]{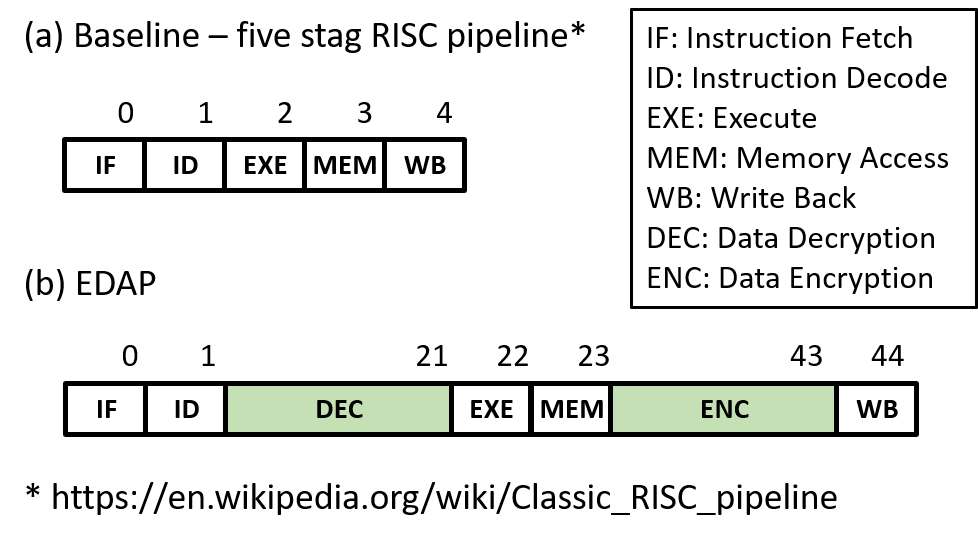}
  \caption{Pipeline structure for (a) the classica five stage RISC design and (b) the EDAP implementation}
  \label{fig:baselinepipeline}
\end{figure}

Decrypting and encrypting data for each instruction adds cycle time to the overall execution and can be illustrated by comparing the pipeline structure of baseline design with this implementation.  In a classic five stage RISC pipeline, there are instruction fetch (IF), instruction decode (ID), execute (EX), memory access (MEM), and register write back (WB) as shown in Figure \ref{fig:baselinepipeline}.  Since the data are encrypted throughout the memory system, including register file, in this EDAP implementation, the five state RISC pipeline need to be extended with additional two stages to decrypt (DEC) the register value before the execute stage and to encrypt (ENC) the result before the register write back as shown in Figure \ref{fig:baselinepipeline}.  The encryption and decryption overhead are typically in the order of tens of processor cycles.   

\begin{figure}
  \includegraphics[scale=.5]{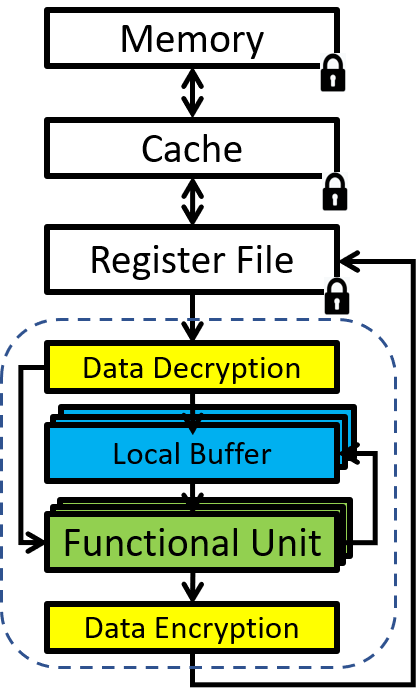}
  \caption{EDAP implementation with data encryption engine placed inside the function unit encalve and the cleartex local buffer to cache recently referenced data}
  \label{fig:EDAPimplementation2}
\end{figure}

To reduce the decryption and encryption overhead, a local buffer can be added to each of the functional units to store the cleartext of its recently accessed registers to allow the same application to bypass the data decryption of these cache value as shown in Figure \ref{fig:EDAPimplementation2}.  The data remains encrypted in the register file with the cleartext of a subset of the registers stored in the buffer that is private to each functional unit.  To protect the cleartext data, each entry of the buffer is accessible to only the authorized application in the user mode and not to other software.  The cleartext data is erased when the thread is switched out of the execution pipe.  Again, this scheme is completely transparent to the software and the data is protected throughout the system.

\begin{figure}
  \includegraphics[scale=.5]{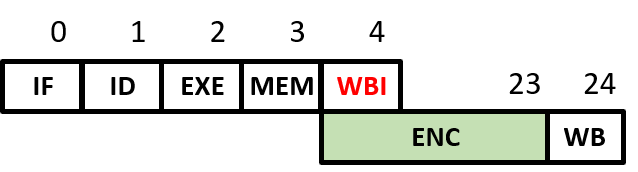}
  \caption{Pipeline structure for EDAP implementation with cleartext local buffer}
  \label{fig:EDAPpipeline2}
\end{figure}

\begin{figure}
  \includegraphics[scale=.4]{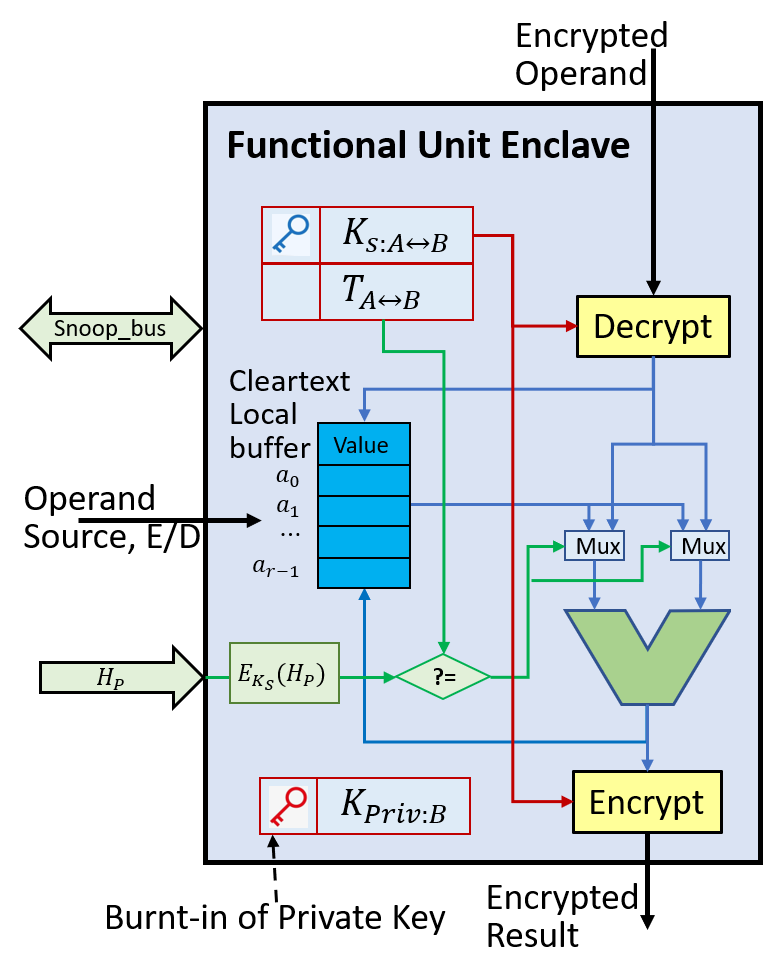}
  \caption{Using snoop bus to clear cleartext in the local buffer}
  \label{fig:EDAPsnoopclear}
\end{figure}

If the source data of an instruction is found in the buffer, the cleartext data is read from the buffer rather than from the register file and the data decryption stage of the pipeline is bypassed.  The cleartext of all instruction results is stored back to the buffer through the writeback bypass (WBI) while being encrypted (ENC) before written back to register file (WB) as shown in Figure \ref{fig:EDAPpipeline2}.  The buffer resides within the functional unit and is content addressable as a cache for the encrypted register file, tagged with physical register numbers.  Furthermore, there is a snoop bus connected to each buffer to erase the entries when the register is reallocated as shown in Figure \ref{fig:EDAPsnoopclear}.

\subsection{Between L1 Cache and Register File}
\label{cleartextRegfile}
\begin{figure}
  \includegraphics[scale=.35]{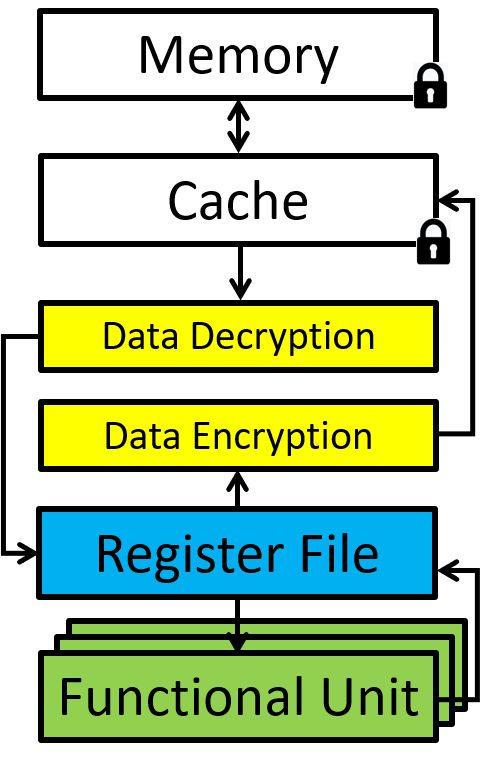}
  \caption{EDAP implementation with data encryption engine placed between L1 cache and register file}
  \label{fig:EDAPimplementation3}
\end{figure}

Placing data encryption engine between L1 cache and register file as shown in Figure \ref{fig:EDAPimplementation3} would lead to a cleartext register file based EDAP implementation.  Each data is decrypted immediately after it is read from the memory to the register file and each data is encrypted right before it is written from the register file back to the memory.  This scheme decouples the data decryption, instruction execution, result encryption, and its writeback to reduce the number of time that a data needs to be decrypted and encrypted in EDAP to alleviate its runtime and power consumption overhead.  To prevent the cleartext data being read by other code or programs throughout the system, the cleartext register file implementation restricts the register access to only its user thread.  Additionally, load and store semantics need to be modified in a way that the data is stored as cleartext in the register file and cyphertext throughout the rest of memory system.  Moreover, since only the authorized application has access to its cleartext registers in problem state, new supporting instructions are needed for the privilege programs, such as hypervisor, supervisor, and interrupt handlers, to save and to restore its program states.  Lastly, the cleartext registers are erased at context switch or when the data is no longer needed.  

\begin{figure}
  \includegraphics[scale=.5]{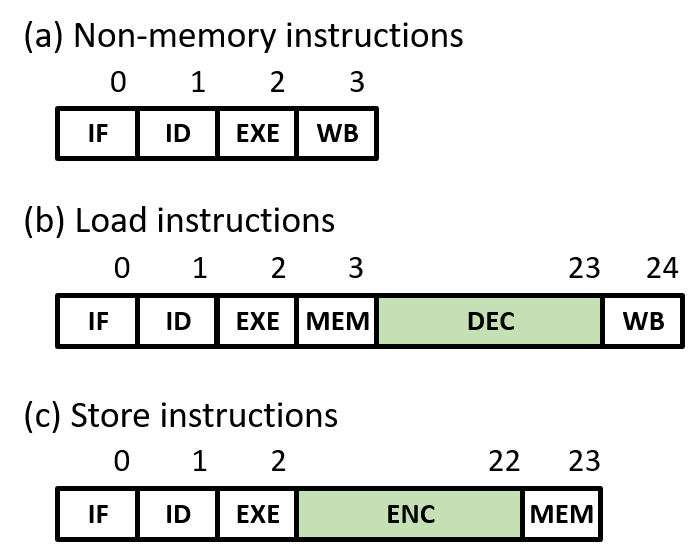}
  \caption{Pipeline structure for (a) non-memory instructions, (b) load instructions, and (c) store instructions in EDAP implementation with cleartext register file}
  \label{fig:EDAPpipeline3}
\end{figure}

Figure \ref{fig:EDAPpipeline3} shows the pipeline for non-memory instructions, load instructions, and store instructions for the EDAP architecture with cleartext register file implementation.  There are no extra cycles added for executing non-memory instructions because the data is already stored as cleartext in the register file as the baseline architecture.  After the encrypted data load from the memory, it is decrypted in the DEC stage to produce cleartext for the register file.  Similarly, the cleartext data is encrypted in the ENC stage before storing it to the memory.  The short pipeline of non-memory instructions is the advantage in placing the data encryption engine between L1 cache and register file.  

In this EDAP implementation, only the user application can access its cleartext registers and this prevents revealing the cleartext data to other programs or code.  Two new privilege instructions, \textbf{\textit{Load-and-Hide}} and \textbf{\textit{Store-and-Clear}}, are added to enable other codes, such as hypervisor, supervisor, and interrupts handlers, to save and restore the state of user application as needed.  The \textbf{\textit{Load-and-Hide}} instruction restores the thread state by first loading the encrypted data from the memory, decrypting the data into cleartext, storing it into the register and hiding it from all codes, including itself, that are not the designated user thread.  The \textbf{\textit{Store-and-Clear}} instruction saves the thread state by first encrypting the cleartext register data, storing the cyphertext to the memory, and clearing the cleartext register from the register file.  These two instructions provide a safe way for other codes to interact with the user application without exposing its data and its encryption keys.   

\subsection{Between L2 Cache and L1 Cache}
\begin{figure}
  \includegraphics[scale=.35]{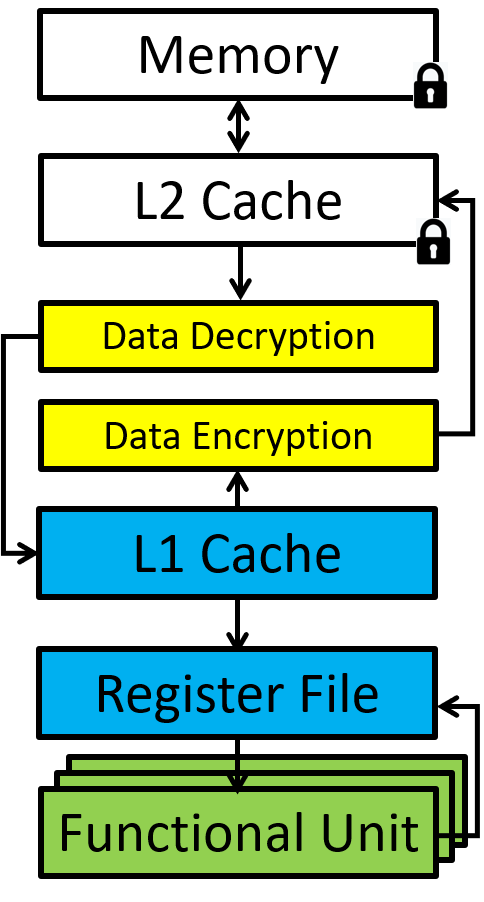}
  \caption{EDAP implementation with data encryption engine placed between L2 cache and L1 cache}
  \label{fig:EDAPimplementation4}
\end{figure}

Moving the data encryption engine to be between L2 cache and L1 cache as shown in Figure \ref{fig:EDAPimplementation4} can further reduce the data encryption and decryption overhead of EDAP architecture.  The L1 cache and register file contain cleartext data and its access is guarded from unauthorized software to protect data confidentiality.  Only the authorized application in problem state can access its cleartext data, like the cleartext register file based implementation in Section \ref{cleartextRegfile}.  When the load and store accesses are L1 cache hit, the instructions skip the data encryption engine.  The extra decryption stage (DEC) is added to load instruction as shown in Figure \ref{fig:EDAPpipeline3} only when the data is not found in L1 cache.  The store instructions store the cleartext data back to L1 cache without the encryption stage (ENC) shown in Figure \ref{fig:EDAPpipeline3} and encrypt the data in parallel to storing it back to the rest of memory system.   

Again, only the authorized application in problem state can read its cleartext data from the register file and L1 cache in this EDAP implementation.  New instructions are added for other software to perform the required supporting tasks without exposing the cleartext data.  In addition to the two privilege support instructions, \textbf{\textit{Load-and-Hide}} and \textbf{\textit{Store-and-Clear}}, described in Section \ref{cleartextRegfile} to enable other codes to save and restore the register state of user application, this implementation calls for a new set of supporting instructions to initialize memory state.  For example, privilege instructions that initialize and set empty data cache block without violating the access permission in EDAP are required.  These instructions are used by supervisor and hypervisor to initialize a block of memory for the authorized application or to create block of memory that is empty.  Moreover, new user instructions for the authorized application to acquire and to release data cache blocks are needed.  These instructions are used by the authorized program in the problem state to initialize an empty block of memory to claim for its private usage or to erase a block of its memory to return it to others to use.


\section{Use Case}
\label{UsageScenarios}
We will describe an EDAP implementation and its use case in detail.  The data encryption engine is placed between L2 cache and L1 cache with XTS-AES memory encryption.  The L1 caches store cleartext data and are indexed by the effective address instead of real address.  The rest of memory system is real addressed and its data is encrypted.  Effective addressed L1 caches simplify the access control and data encryption as each application has its own effective address space and the effective address is part of the encryption scheme.  

The three main actors in the EDAP protocol, the data owner \textbf{DO}, the author of the program \textbf{AA},and the platform provider \textbf{PP}, are described 
in \ref{EDAP_Principals}.
The DO does not trust PP.  In particular, she does not trust any software running on PP's system.  However, DO trusts the manufacturer of EDAP-enabled systems that PP owns.   DO also trusts that AA's program does not leak her data when running, because she either has verified the code herself, trusts AA, or trusts a certification authority that has approved AA's code as safe.

PP owns the data processing equipment and he has the right to deny service to DO.  We are not going to prevent denial of service from PP to DO.  There is a mechanism for PP to prove that DO's program got a certain amount of processing time.  DO can demand that proof before paying PP for his services.  

\subsection{ACT I: DO acquires resources to process her data}
DO contacts PP and asks for an EDAP-enabled machine to process her data.  She can specify number of cores, speed, memory size, etc.  PP agrees to assign an EDAP-enabled machine to DO and gives her the following two items.
\begin{enumerate}
\item
\textbf{\textit{A secure processing identifier} (SEID) $\mathbb{S}$}: A 64-bit tag that is to be used just by DO, if PP is well-intentioned.  If PP is ill-intentioned, it can be anything he wants, but it will not compromise DO's data in any way.  
\item
\textbf{\textit{A processor public key} (PPK) $\mathbb{P}$}: A bit-string that is part of the public/private key pair of an EDAP-enabled machine that its processor manufacturer assigned.  This public key is published by the manufacturer and DO can verify that a particular public key exists.
\end{enumerate}
At this point, DO can safely assume that an EDAP-enabled processor is ready to process her data.  If that is not the case, her data will still be safe.  

\subsection{ACT II: DO prepares her program}
\begin{figure}
  \includegraphics[scale=.4]{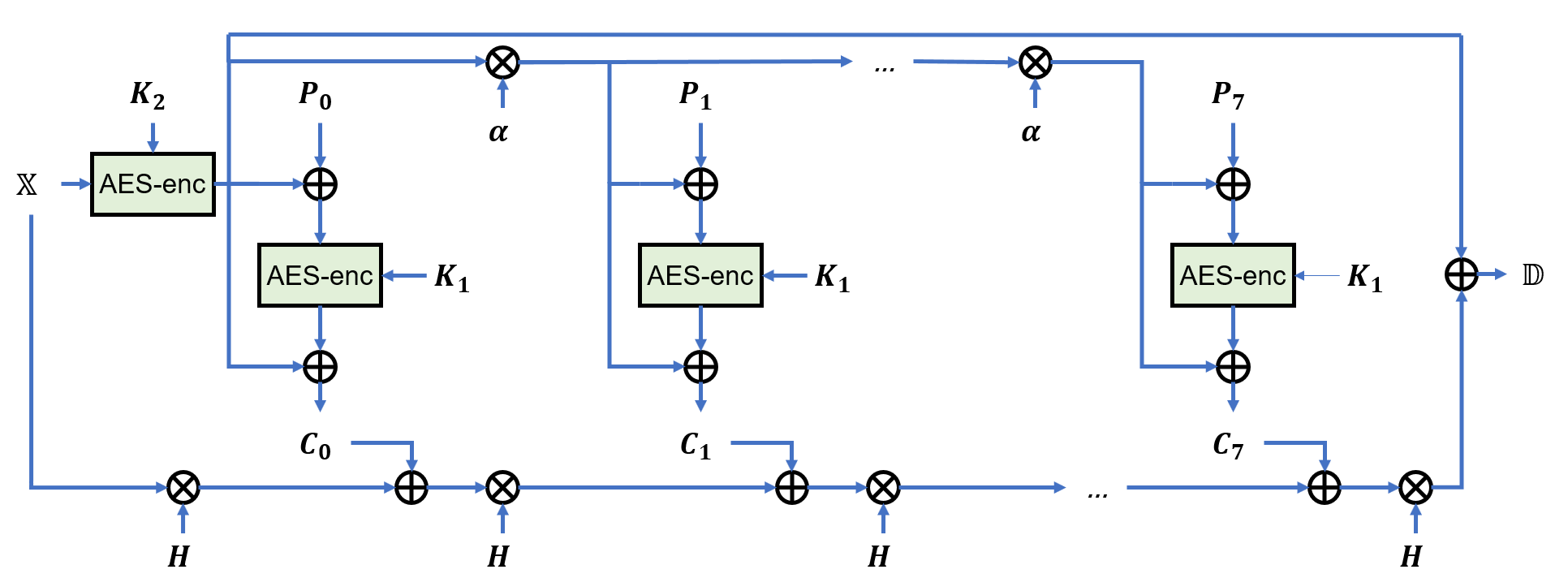}
	\caption{XTS-AES encryption combined with AES-GCM authentication. Each 128-byte cache line block is chopped into eight 16-byte blocks and ecnrypted according to the XTS-AES scheme (top part of drawing). As the sections are encrypted, a digital signature is computed using AES-GCM scheme (bottom part of drawing.)}
  \label{fig:XTSAESEncryption}
\end{figure}

DO selects a secret XTS-AES Key $\mathbb{K}=\langle K_1, K_2 \rangle$.  This pair of either 128- or 256-bit encryption keys is known only to DO.  Using this key, DO prepares a secure executable, staring form the executable file of AA's program.  Every code and data section of the executable consists of multiple naturally aligned 128-byte blocks, the size of a cache line.  As shown in Figure \ref{fig:XTSAESEncryption}, each 128-byte block $\Gamma$ is XTS-AES encrypted using $\mathbb{K}$ and tweak $\mathbb{X}=\langle \mathbb{S}, \mathbb{E} \rangle$, where $\mathbb{E}$ is the effective address of the block, to produce a 128-byte ciphertext, $\Delta$.  The 128-byte $\Gamma$ and $\Delta$ blocks are evenly chopped into eight 16-byte sections, $\Gamma=\langle P_0, P_1, ..., P_7 \rangle$ and $\Delta=\langle C_0, C_1, ..., C_7 \rangle$.  $\alpha$ is a primitive element of $GF(2^{128})$, chosen as the polynomial $x$.
In parallel with the XTS-AES encryption, we use the AES-GCM scheme to cryptographically sign each 128-byte block with a 8-byte digest $\mathbb{D}$, so it cannot be tampered or relocated. $H = E_{K_2}(0)$ is the hash key for the AES-GCM scheme. SSL credentials can be built into the executable and are independent of the memory encryption scheme.  Secret data can also be part of the executable, as static data.  
 
\subsection{ACT III: DO packages and sends her code to PP}
DO creates a secure executable that includes all necessary problem-state code as she does not trust any software in PP's systems, including installed libraries.  Next, DO generate a \textit{session key} $\mathfrak{K}$ and establishes a secure session with PP's EDAP-enabled processor by sending $E_{\mathbb{P}}(\mathfrak{K})$, the session key encrypted with public-key $\mathbb{P}$.  At this point, DO can send the executable to PP as it will be opaque to PP.  The executable is sent through a streaming protocol (e.g., AES-GCM) using session key $\mathfrak{K}$ as sequence of $\langle \mathbb{E}, \mathbb{C}, \mathbb{D}\rangle$ tuples.  $\mathbb{E}$ is the effective address, $\mathbb{C}$ is the 128-byte block ciphertext, and $\mathbb{D}$ is the 8-byte digest of the ciphertext block.  A streaming protocol is used to send the executable from DO to PP to ensure the processor does not accept the same tuple twice.  Finally, DO sends PP $E_{\mathbb{P}}(\mathbb{K})$, the XTS-AES key pair $\mathbb{K}$ encrypted with public-key $\mathbb{P}$.  Also, DO should never use the same $\langle \mathbb{K}, \mathbb{S}\rangle$ pair again to prevent replay attacks.   

\subsection{ACT IV: PP loads and runs DO's code}
PP creates a process to run DO's code.  Each 128-byte block of DO's program is loaded raw (ciphertext and its digest) in the real address that corresponds to the effective address of the block.  This must be done through the secure session that DO has with the EDAP-enabled processor to prevent replay after execution starts.  The load is done with a new supervisor/hypervisor operation that initializes a block of empty real memory.  If either the ciphertext or the digest are tampered with, code will not run.  Additionally, if a block is mapped to different effective address, the code will not run.  Code can only be run by the intended EDAP-enabled processor with XTS-AES key pair $\mathbb{K}$.

PP then loads $E_{\mathbb{P}}(\mathbb{K})$ into the intended EDAP-enabled processor.  The processor uses its private key to extract $K_1$ and $K_2$ that are used in the XTS-AES encryption by DO.  DO's program will not run if the wrong key is loaded.  Next, PP transfers control to DO's program.  This transfer of control must clear the L1 caches, both data and instruction, and engage in encryption engine.  From this point on, only DO's program will run and anything else will fail integrity check. We now have a secure DO's process running on PP's system.  

Any transfer of control into supervisor/hypervisor must clear the L1 caches and disengage the encryption engine.  Any transfer of control back to problem state must clear the L1 caches and reengage the encryption engine.  The clearing of L1 cache can be finer grained than state as only the affected process needs to be impacted.  Additionally, the processor remembers the cryptographic hash of register state to check when resuming execution.  This can be a table, indexed by $\mathbb{S}$, inside the processor or a linked-list in memory.

\subsection{ACT V: DO's process acquires more secrets}
Additional data can be acquired from DO through a variety of ways.  One way is to establish secure sockets and directly stream data from DO to her running process.  This is different than the secure session DO has with PP's EDAP enabled processor.  If successful, it means DO's program is running on the target EDAP-enabled system as no other system would be able to run DO's program.  Another way is for DO to place encrypted files in shared storage and keys can be built into the executable or loaded at run time.  Regardless of the methods, every time a cache block is read or written, integrity must be check.  Violation means either the tweak or the keys are not right and would cause exception.  Also, encryption engine is used to encrypt and sign data before storing to memory.

\subsection{ACT VI: DO's process runs to completion}
DO's process runs for as long as it wants if PP lets it run.  No software other than DO's program, running in DO's process, can access DO's data.  For supervisor/hypervisor to support DO's process, a new set of services is architected.    These services are allocating memory, initializing memory, loading new code, reading/writing files and sockets, ensuring that DO's code always starts from the entry point and can only be started in the correct address space, and various other system calls.  There are a lot of details to be sorted out in the service architecture, but these are all done at discretion of DO's program, so they can be made safe.  The important feature is that supervisor/hypervisor simply cannot meaningfully view/write DO's code or data on their own.  DO's process must explicitly ask for the service.

\section{Performance Evaluation}
\label{Sec:PerformanceEvaluation}
\label{sec:Simulation_EDAP}

We use a simulation model to gain a better understanding of the performance impact of data encryption and data decryption overhead in an EDAP-enabled system.  Our simulation model is a trace driven and cycle accurate timing model of an IBM POWER10 like processor core with four execution pipes.   It has a 48kB L1 instruction cache, a 32kB L1 data cache, and a 1MB L2 cache.  Two of the EDAP implementations in Section \ref{Implementation}, one with the least trusted hardware footprint and one with the most trusted hardware footprint, are modeled in this study.  We first implement the EDAP design where the data encryption engine is placed within the functional unit enclave without the cleartext caching.  Each non-memory instruction would incur 40 processor cycles to decrypt its source data and encrypt its output data as this processor is deeply pipelined.  The data encryption engine is fully pipelined and has the capacity to process one instruction per execution pipe.    

\begin{figure}
  \includegraphics[scale=.8]{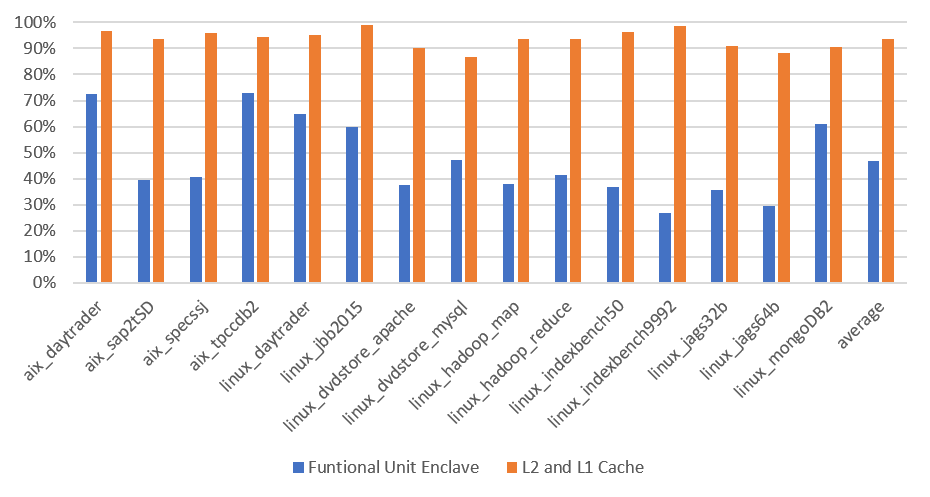}
  \caption{Normalized instruction per cycle (IPC) for two EDAP implementations: [1] data encryption engine inside the functional unit enclave without cleartext buffering (blue bars) and [2] data encryption engine in-between L2 cache and L1 cache of processor core (orange bars).  The results are normalized to non-EDAP version of baseline processor that has 100\% on each workload.}
  \label{fig:EDAPResultCPC}
\end{figure}

We chose a collection of commercial workload traces for our study.  Each trace is 1M instruction long.  Figure \ref{fig:EDAPResultCPC} shows the performance of our EDAP processor core comparing to the baseline processor by using the simulation model in two configurations.  First, the data encryption engine is placed within the functional enclave.   The 40-cycle encrypted data overhead, 20 cycle for decryption and 20 cycle for encryption, is incur in every non-memory instruction.  The simulation model indicates that the EDAP-enabled core runs at a 27\% to 73\% of baseline speed, with an average of 47\% across the workload.  This means that the EDAP-enabled core will take around twice as long as the baseline core to finish executing the same program on average.  



Next, we push the data encryption engine up the memory subsystem to be in-between L2 cache and L1 cache to reduce the EDAP data encryption overhead.  Everything above and including L2 cache remains encrypted.  Both L1 cache and register file now store cleartext data.  For this configuration, the 20-cycle decryption overhead is added only when data is transferred from L2 cache to L1 cache.  If the data is found in the L1 cache in cleartext, the instruction would be executed as in the baseline processor.  However, the 20-cycle data encryption overhead is not added to the store instructions because the encryption is not in the critical path as it can be performed simultaneously to the execution run.  The simulation shows that the performance degradation ranges from 13\% to 1\% with an average of 6\%.  By relocating the data encryption engine up from the functional unit enclave to above L1 caches, the performance impact is reduced from 53\% to 6\% on average.  Moreover, the range is smaller across the workload.  

Encryption/decryption of 128-bit data using standard AES algorithm with 128-bit keys, requires a sequence of 10 rounds of logic.  Our experimental designs to do the rounds directly in optimized logic suggested that each round can be done under 2 cycles in
contemporary technology.  Based on that, our simulations used a penalty of 20 cycles for encryption (same for decryption).  In some cases we also have to encrypt the address; but that can be done in parallel and folded into the pipleline without additional impact.  We have also done simulations to see the sensitivity to this latency, by doubling and halving the latency.  We notice reasonable linear response to these changes.

\section{Conclusions}
\label{sec:Conclusions}
EDAP is a comprehensive architecture for confidential computing without the requirement of trusting platform provider or other software such as supervisor, hypervisor, OS, or other users' applications running on the same platform.  It is general purpose, efficient, and can run the application as is.  Data is fully protected as only the applications that are authorized by data owner have access to its cleartext on the targeted EDAP-enabled machine.  Although our intuition is that the performance penalty of adding data encryption engine to the processor can potentially slowdown the program execution by multiple orders of magnitude, our simulation model indicates that the pipelining and out-of-order execution characteristic of modern processors can alleviate the performance degradation.  The performance impact is sustained within 15\% with an average of 6\% across the commercial workload that we simulated when the data encryption engine placed above the L1 cache.  It is increased to 53\% on average if the data encryption engine is moved inside the functional unit enclave to remove the need of providing new set of processor operations that enable system software such as hypervisor or supervisor to perform supporting operations to the application without exposing it's data.  In this paper, we demonstrate that EDAP is a suitable solution to confidential computing in terms of its ease of adoption and the overall performance.  An interesting area of future research is deriving the contractual agreement for security of data/information using the principal-agent problem framework \cite{PrincipalAgent}.

\bibliographystyle{ACM-Reference-Format}
\bibliography{EDAP}

\end{document}